\documentclass[
    ,final            
  ,numberedheadings 
  ]{aipproc}

\layoutstyle{8x11single}
\usepackage{amsmath,amssymb}
\usepackage{./floatflt}
\usepackage[bf]{caption}
\renewcommand{\figurename}{Fig.}

\begin{document}

\title{Kochen-Specker Algorithms for Qunits}

\author{Mladen Pavi\v ci\'c\footnote{E-mail: pavicic@grad.hr;  
Web page: \tt{http://m3k.grad.hr/pavicic}}}{
address={University of Zagreb, Zagreb, Croatia}}

\begin{abstract}
Algorithms for finding arbitrary sets of Kochen-Specker (KS) 
qunits ($n$-level systems) as well as all the remaining vectors 
in a space of an arbitrary dimension are presented. The algorithms 
are based on linear MMP diagrams which generate orthogonalities of 
KS qunits, on an algebraic definition of states on the diagrams, and on
nonlinear equations corresponding to MMP diagrams whose solutions 
are either KS qunits or the remaining vectors of a chosen space 
depending on whether the diagrams allow {\tt 0-1} states or not.
The complexity of the algorithms is polynomial. New results 
obtained with the help of the algorithms are presented. 
\end{abstract}

\maketitle

\section{Introduction}
In quantum measurements, particular directions of quantisation 
axes of the measured observable obey the following KS conditions: 
(1) no two of mutually orthogonal directions can both be assigned 1; 
(2) they cannot all be assigned 0.~\cite{zimba-penrose}
To these directions we ascribe vectors that we call {\em KS vectors}.
To stress that the axes and therefore KS vectors themselves cannot 
be given a {\tt 0-1} valuation, i.e.~represented by classical bits, 
we also call them {\em KS qunits} ({\em qu}antum {\em n}-ary 
dig{\em its}). We are not aiming at giving yet another proof of 
the KS theorem but at determining the 
class of all KS vectors from an arbitrary ${\cal H}^n$ as well as 
the class of all the remaining vectors from ${\cal H}^n$. In other 
words, our aim is to obtain all physically realisable vectors in 
${\cal H}^n$ corresponding to orientations of projectors. 

In order to achieve this aim, we first recognise that a description 
of a discrete observable measurement (e.g., spin) in ${\cal H}^n$ 
can be rendered as a {\tt 0-1} measurement of the corresponding 
projectors along orthogonal vectors in ${\mathbb R}^n$ to which 
the projectors project. Hence, we deal with orthogonal triples 
in ${\mathbb R}^3$, quadruples in ${\mathbb R}^4$, etc., which 
correspond to possible experimental designs. To find KS vectors 
means finding all such $n$-tuples in ${\mathbb R}^n$ that correspond
to experiments which have no classical counterparts.  

\section{MMP diagrams and their generation}

We represent the $n$-tuples in ${\mathbb R}^n$ by means of 
points of a diagram. The points are called {\em vertices}. Groups of 
orthogonal vectors are represented by connected vertices and are 
called {\em edges}. Vertices and edges form linear {\em MMP 
diagrams} \cite{bdm-ndm-mp-1,mporl02} which are defined as follows:

1.~Every vertex belongs to at least one edge;

2.~Every edge contains at least 3 vertices;

3.~Every edge which intersects with another edge
at least twice contains at least 4 vertices. 

The procedure for generation MMP diagrams follows from 
isomorphism-free generation of linear diagrams established in 
\cite{bdm-ndm-mp-1,mporl02,mckay98}. Here we  present 
the main idea behind such a generation and will present the
details elsewhere.~\cite{pmmm04a-arXiv}

The algorithm for obtaining MMP diagrams is based 
on the parent-child generation tree as the one shown in Fig.~1
for the particular number of vertices and loops we allow
(it will be shown below that for 3 vertices per edge no 
diagram containing loops of size less then 5 can generate a 
set of vectors). 
At every further level we add an edge until we reach a 
desired number of vertices and edges. In doing so the main 
problem is to avoid generation of diagrams isomorphic to an 
already generated diagram. This is resolved by using the 
procedure given below. The procedure guarantees 
that at each parent level generation of isomorphic 
children is suppressed. E.g., each of the two diagrams at the 
third level could give the third diagram at the fourth level. 
However, the generation procedure prevents the second generation.    
 
\begin{floatingfigure}{0.53\textwidth}
\begin{center}
\includegraphics[width=0.53\textwidth]{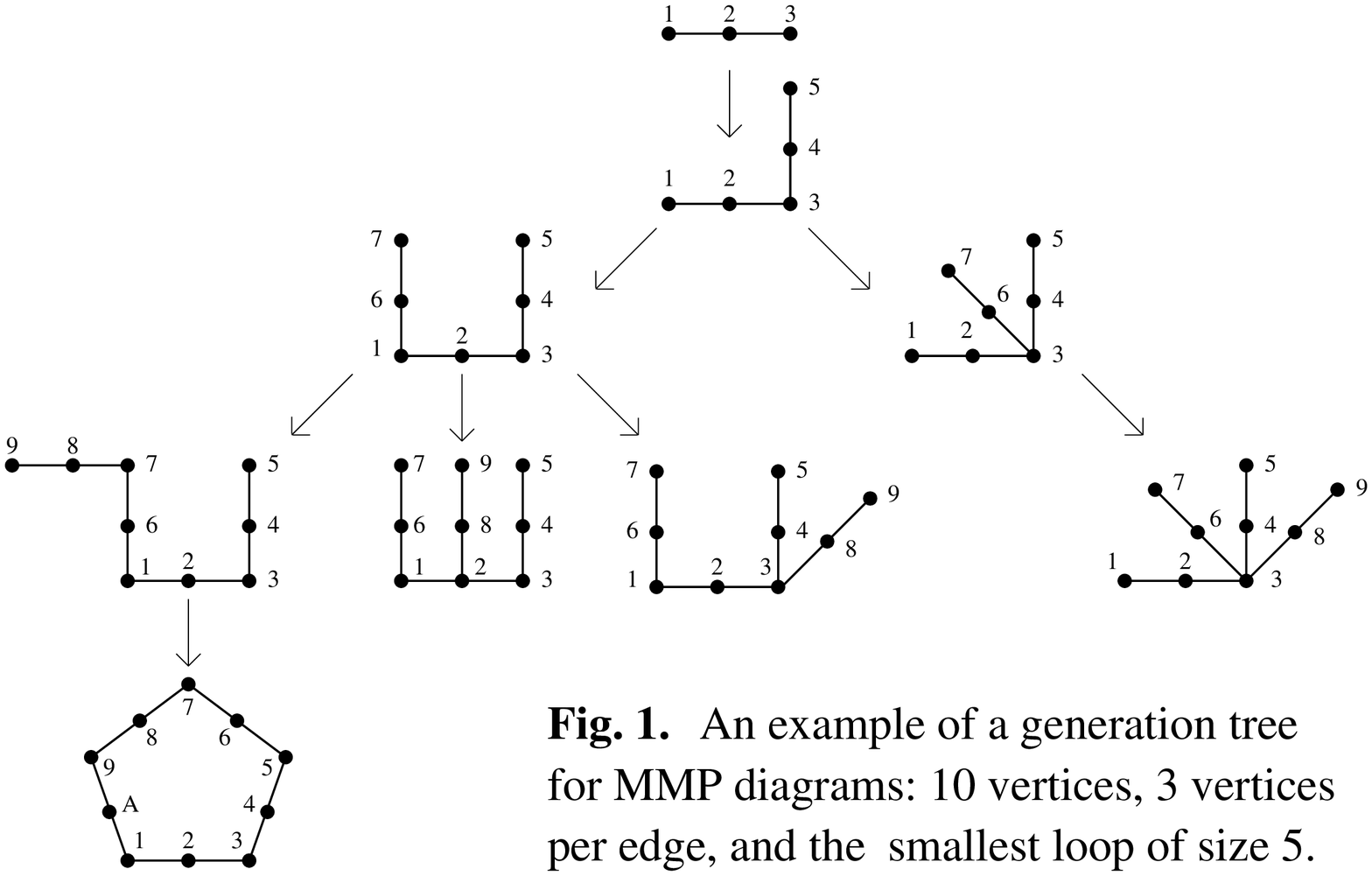}
\end{center}
\end{floatingfigure}

\noindent
{\bf procedure} scan $(D$ {\rm : diagram;} $\beta$ {\rm : integer}$)$

{\bf if} $D$ {\rm has exactly $\beta$ edges} {\bf then}

\quad{\bf output $D$}

\noindent
{\bf else}

{\bf for} {\rm each equivalence class of extensions} $D+e$ {\bf do}

\quad {\bf if} $e\in m(D+e)$ {\bf then} scan($D+e$,$\beta$)

\noindent
{\bf end procedure}

It is the function $m(\cdot)$ which we define in 
\cite{bdm-ndm-mp-1} that enables us to define a unique 
isomorphism class as the parent class of the isomorphism 
class of $D$. This way of generation of diagrams enables 
a very efficient parallelisation of computing on clusters. 
The complexity of generation grows exponentially and if we
had not integrated its algorithm with the related algorithms 
for solving the corresponding 
non-linear equations we would not have been able to 
realistically implement the generation of MMP diagrams. 
Because, e.g., there are 6 3-vertices-per-edge and smallest 
loops of size 5 diagrams with 19 vertices and 13 edges 
and already 7447274324 diagrams with 29 vertices and 20 edges, etc.

We denote vertices of MMP diagrams by {\tt 1,2,..,A,B,..a,b,..}
By the above algorithm we generate MMP diagrams with
chosen numbers of vertices and edges and a chosen minimal loop size.

\section{0-1 states on MMP diagrams}

\begin{floatingfigure}{0.375\textwidth}
\begin{center}
\includegraphics[width=0.375\textwidth]{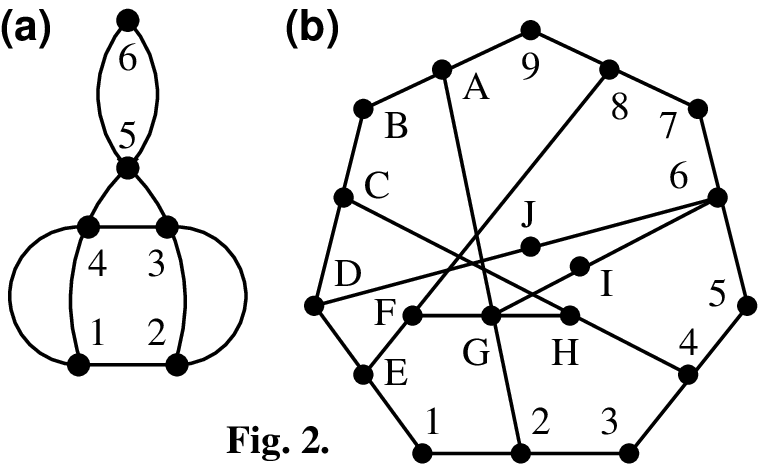}
\end{center}
\end{floatingfigure}

To find diagrams that do not allow {\tt 0-1} states we use 
an algorithm which carries exhaustive search on MMP 
diagrams with backtracking. The idea of the algorithm 
follows from our elaboration on states in the Hilbert 
space.~\cite{mpoa99} Essentially, the algorithm 
ascribes non-dispersive ({\tt 0-1}) states so as to assign 
1 to exactly one vertex within an edge while the other 
vertices within the edge are assigned 0. The algorithm 
scans the vertices in some order, trying 0 then 1,
skipping vertices constrained by an earlier assignment.  When no
assignment becomes possible, the algorithm backtracks until 
all possible assignments are exhausted or a valid assignment is found.
Again, the algorithm by itself would be exponential, 
but the constraints build up quickly so that the  
exponential behaviour of the algorithm is reduced to a polynomial one.

In Fig.~2 (a) the smallest diagram with 4 vertices per edge and 
with the smallest loop of size 2 (two edges share two vertices) 
which does not admit {\tt 0-1} states is shown. It has 6 vertices 
and 3 edges, so it is denoted as 6-3. We can write it down as 
follows: {\tt 1234,2356,1456.} Fig.~2 (b) shows one of the two 
smallest diagrams with 3 vertices per edge and smallest loops of 
size 5 that do not admit {\tt 0-1} states: 19-13: 
{\tt 123,345,567,789,9AB,BCD,DE1,2GA,4HC,6IG,6JD,8FE,FGH.}

\section{Kochen-Specker Qunits}

To obtain KS qunits we merge the algorithms for generation of 
linear MMP diagrams corresponding to edges of orthogonal vectors 
in ${\mathbb R}^n$ (Sec.~2), algorithms for 
filtering out those vectors on which {\tt 0-1} 
states cannot be defined (Sec.~3), and algorithms for  
solving nonlinear equations describing the orthogonalities 
of the vectors by means of interval analysis so as to eventually 
generate KS vectors in a polynomially complex way.

Following the idea put forward in \cite{mporl02} we proceed so 
as to require that the vertices correspond to vectors from 
${\mathbb R}^n$, that the number of vertices within edges
corresponds to the dimension of ${\mathbb R}^n$, and that edges 
correspond to $n(n-1)/2$ equations resulting from inner products 
of vectors being equal to zero which means orthogonality.
An example for the edge {\tt BCD} in ${\mathbb R}^3$ is shown 
by Eq.~(1). 

Combinations of edges for a chosen number of vertices and 
chosen number of orthogonalities between them 
correspond to a system of such nonlinear equations. 
\begin{floatingfigure}{0.45\textwidth}
\begin{center}
\vskip-20pt
\begin{eqnarray}
&{\mathbf a}_B\cdot{\mathbf a}_C=
a_{B1}a_{C1}+a_{B2}a_{C2}+a_{B3}a_{C3}=0,&\nonumber\\
&{\mathbf a}_B\cdot{\mathbf a}_D=
a_{B1}a_{D1}+a_{B2}a_{D2}+a_{B3}a_{D3}=0,&\nonumber\\
&{\mathbf a}_C\cdot{\mathbf a}_D=
a_{C1}a_{D1}+a_{C2}a_{D2}+a_{C3}a_{D3}=0.&
\end{eqnarray}
\vskip-15pt
\end{center}
\end{floatingfigure}

However, we cannot first generate all possible MMP diagrams 
and only afterwards try to filter them by solving equations because
their number grows exponentially and the required time for 
such a generation on the fastest CPU would exceed the age of 
the universe for any application of interest. Instead, we restrict 
generation of MMP diagrams so that as soon as a preliminary pass 
determines that an initial set of $m$ $n$-tuples has no solution, 
no further systems containing this set will be generated and the 
whole sub-tree in the generation tree is discarded. 
E.g., for 18 vertices and 12 quadruple edges without 
such a restriction one should generate $> 2.9\cdot 10^{16}$ 
systems---what would require more than 30 million years on a 2 GHz 
CPU---while the filter reduces the generation to 100220 systems 
(obtainable within $<30$ mins on a 2 GHz CPU). Thereafter 
we get 26800 systems without {\tt 0-1} states in $<5$ secs. 
We also developed a checking program that finds solutions from
assumed sets, say $\{-1,0,1\}$, even much faster ($<$ 1 sec on a
2 GHz CPU).

For the remaining systems two solvers have been developed. One is
based on a specific implementation of Ritt characteristic set
calculation and the other on the interval analysis, in particular
the library 
ALIAS\footnote{\url {www.inria.fr/coprin/logiciels/ALIAS/ALIAS.html}}. 
We will elaborate on these solvers and present further details of the 
algorithms elsewhere.~\cite{pmmm04a-arXiv} 

\section{Some Results}

Using our general algorithms presented above, we wrote 
programs~\cite{pmmm04a-arXiv} which provided us with a number of 
special results that show the power of the algorithms.

(i) A general feature we found to hold for all MMP diagrams without 
{\tt 0-1} states we tested is that the number of vertices $a$ and 
the number of edges $b$ satisfy the following inequality: 
$nb\ge2a$, where $n$ is the number of vertices per edge. 
In ${\mathbb R}^n$ this means that we cannot arrive at systems 
with more unknown (components of vectors) then equations 
(orthogonality conditions). The biggest systems we tested 
contain 30 vertices. There is only one system known to 
us---containing 192 vertices---for which the inequality is 
violated.~\cite{pmmm04a-arXiv}  

(ii) Several self-explaining results are presented in Fig.~3.
As we can see, two of them have miraculously been previously 
found by some ingenious {\em humans} (think of billions of 
positions the vectors can take). 

\setcounter{figure}{2}

\begin{floatingfigure}{\textwidth}
\begin{center}
\vskip-155pt
\includegraphics[width=\textwidth]{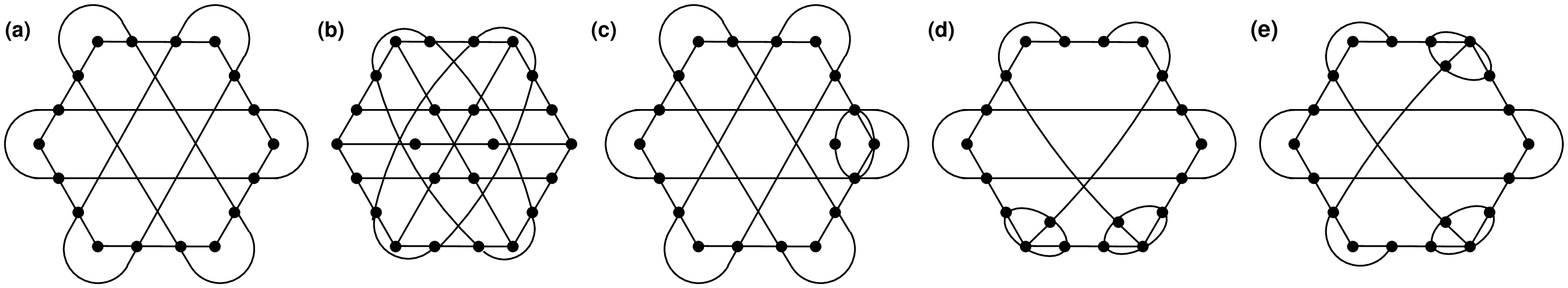}
\renewcommand{\figurename}{Fig.}
\caption{\it Smallest 4-dim KS vectors with: 
\ {\bf (1)} \ loops of size 3: (a) 18-9
(isomorphic to Cabello et al.~{\rm \cite{cabell-est-96a}}); (b) 24(22)-13
not containing (a), with values $\not\in\{-1,0,1\}$; \ {\bf (2)} \ 
loops of size 2: {\bf (2.1)} containing (a): (c)~19(18)-10; 
{\bf (2.2)} not containing (a): (d)~20-11-a; (e)~20-11-b
(isomorphic to Kernaghan~{\rm \cite{kern}}). Solutions are given in 
{\rm \cite{pmmm04a-arXiv}}.}
\end{center}
\end{floatingfigure}

\vskip155pt
\ 

(iii) Between 4-dim systems, with smallest loops of size 3,  
(a) and (b) of Fig.~3, there are 62
systems with loops of size 3, all containing the system (a), 
but 37 of whom do not have solutions from $\{-1,0,1\}$. 
System (b) is the first system not containing (a); it does 
not have a solution from  $\{-1,0,1\}$. The solution is 
given in ~\cite{pmmm04a-arXiv}.

(iv) All 4-dim systems with up to 22 vectors and 12 edges with 
the smallest loops of size 2 which do have solutions from  
$\{-1,0,1\}$ contain at least one of the systems (d) and (e) and 
in many cases also (a) of Fig.~3. The two smallest 4-dim systems
with the smallest loops of size 2 that contain neither of 
the latter three systems are 22-13 systems (a) and (b) of 
Fig.~4: 
{\tt 1234,4567,789A,ABCD,DEFG,GHI1,2ILA,345J,4JEC,678K,7KMG,9ABL,FGHM}
and 
{\tt 1234,4567,789A,ABCD,DEFG,GHI1,12IJ,345K,678L,GML7,1J9B,4KEC,FGHM}.

(v) \ \ \ (c) and (d) of Fig. 4 are two presentations of the same 24-24 
KS system: {\tt 1234,4567,789A,ABCD,DEFG,
GHI1,\hfil 12IJ,\hfil 345K,\hfil 678L,\hfil 7LOG,\hfil 68FH,\hfil 1J9B,\hfil AMI2,\hfil 4KCE,\hfil DN35,\hfil CDEN,\hfil IJK5,\hfil 38KL,\hfil 6BML,\hfil 9EMN,\break 
CHNO,2JOF,9ABM,OFGH} that 
make it easier to recognise that the system contains all the previous 
systems: (a), (c), (d), and (e) of Fig.~3 and (a) and (b) of Fig.~4. 
The vectors determined by the solution of the system are nothing 
but Peres' vectors~\cite{peres}, although he has most probably 
never tried to identify all 24 tetrads for his 24 vectors 
(our programs verify all these options in a fraction of a second).  
We also developed a program for drawing MMP diagrams.~\cite{mpoa99} 

\vskip-10pt
\begin{floatingfigure}{\textwidth}
\begin{center}
\vskip-190pt
\includegraphics[width=\textwidth]{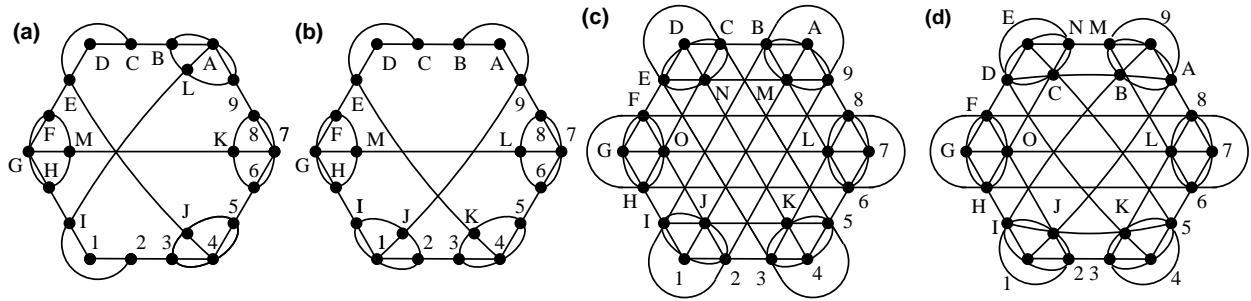}
\renewcommand{\figurename}{Fig.}
\vskip-14pt
\caption{\it (a) and (b): Smallest KS systems systems (22-13) 
containing neither (a) nor (c) nor (d) nor (e) of Fig.~3. Solutions are given 
in {\rm\cite{pmmm04a-arXiv}}; (c) and (d) (isomorphic to each other)
are a 24-24 KS system containing (a), (c), (d), and (e) of 
Fig.~3 and all Peres' vectors {\rm\cite{peres}}.} 
\end{center}
\end{floatingfigure}

\vskip185pt
 
\ 

(v)~4-dim sys\-tems with more than 41 vectors cannot have solutions
from $\{-1,0,1\}$ and there are no such solutions to
systems with minimal loops of size 5 up to 41 vectors, what brings the
Hasse (Greechie) diagram approach to the KS problem into question.
Up to 41 vertices with loops of size 5 there are only two 
diagrams that have no 0-1 states.

(vi) It can easily be shown that a 3-dim system of equations 
representing diagrams containing loops of size 3 and 4 
cannot have a real solution.~\cite{pmmm04a-arXiv} 

(vii) We scanned all systems with up to 30 vectors
and 20 orthogonal triads and there are no KS vectors among them.
This does not mean that Conway-Kochen's system is the smallest
KS system, though.~\cite{pmmm04a-arXiv} It turns out that we cannot drop 
vectors that belong to only one edge from orthogonal triads because 
there are cases where a solution to a full system allows 
{\tt 0-1} valuation while one to a system with dropped vectors 
does not or where the full system cannot 
have a solution. So, Conway-Kochen's system is actually not a 
31 but a 51 vector system.~\cite{pmmm04a-arXiv} See also \cite{larsson}.

(viii) The class of all remaining (non-KS) vectors from ${\cal H}^n$
we get so as to first filter out all the MMP diagrams that do 
have {\tt 0-1} states. Then the real solutions to the equations 
corresponding to these diagrams yield the desired vectors. 

(ix) Presented algorithms can easily be generalised beyond the 
KS theorem. One can use MMP diagrams to generate Hilbert
lattices, partial Boolean algebras, and general quantum algebras 
which could eventually serve as an algebra for quantum 
computers.~\cite{mpoa99} One can also treat
any condition imposed upon inner products in ${\mathbb R}^n$ to 
find solutions not by directly solving all nonlinear equations but
by first filtering the corresponding diagrams and solving only
those equations that pass the filters.




\end{document}